\documentclass[12pt,english]{article}
\usepackage[T1]{fontenc}
\usepackage[latin9]{inputenc}
\usepackage{amsmath}
\usepackage{amssymb}

\makeatletter

\usepackage{amsthm}\usepackage{amsfonts}\usepackage{epsfig}\usepackage{graphics}\@ifundefined{definecolor}
 {\usepackage{color}}{}

\theoremstyle{definition}

\theoremstyle{remark}

\numberwithin{equation}{section}

\makeatother

\usepackage{babel}

\begin{document}

\begin{center}

{\Large {\bf The Energy Transformation Limit Theorem for Gas Flow Systems} }\\ [5mm]

{\bf V.T.\,Volov} \\ [3mm]

{\it Physical Department, Samara State University of Railway Transport,  Samara, Russia\\
Samara Institute of Fundamental Research, Samara, Russia\\
e-mail: vtvolov@mail.ru} \\ [2mm]
\end{center}

\begin{abstract}
The limit energy theorem which determines the possibility of transformation
the energy flow in power systems in the absence of technical work
is investigated and proved for such systems as gas lasers and plasmatrons,
chemical gas reactors, vortex tubes, gas-acoustic and other systems,
as well as a system of close stars. In the case of the same name ideal
gas in the system the maximum ratio of energy conversion effectiveness
is linked to the Carnot theorem, which in its turn is connected with
the Nernst theorem. However, numerical analyses show that the class
of flow energy systems is non-carnot one. The ratio of energy conversion
effectiveness depends on the properties of the working medium; a conventional
cycle in open-circuit is essentially irreversible. The proved theorem
gives a more strongly worded II law of thermodynamics for the selected
class of flow energy systems. Implications for astrophysical thermodynamic
systems and the theory of a strong shock wave are discussed.
\end{abstract}

\section{Introduction}

There is a wide class of flow energy systems (FES) with a fast-flow
gas where technical work is absent ($L_{tech}=0)$. These energy systems
include: gas lasers and plasmatrons; chemical gas reactors, vortex
tubes, ejectors, mixers, acoustic devices, and some astrophysical
objects, such as systems of close stars.

The common property of these systems is that their energy efficiency
is higher, the higher is the degree of transformation of kinetic energy
of the flow into potential energy of pressure at the outlet of FES.
However, there was no assessment of the maximum efficiency of energy
transformation in the given class of energy systems in scientific
literature so far. And as a rule, assessment of the effectiveness
of these systems was carried out when compared with the Carnot cycle.

To determine the maximum efficiency of energy transformation in FES
we will prove the following theorem.

\section{The theorem}

The efficiency ratio of energy transformation $\eta_{G>0}^{ideal}$
in an ideal flow energy system can not exceed the value
\begin{equation}
\Delta\overline{N}_{max}=\dfrac{\displaystyle{\sum\limits _{i=1}^{m}}\dfrac{\gamma_{1}-1}{\gamma_{1}}\mu_{i}\overline{R}_{i}\Theta_{i}\left(1+\dfrac{\lambda_{i}^{\text{2}}}{\gamma_{\text{i}}+1}\right)+
\overline{\dot{Q}}}{\displaystyle{\sum\limits _{i=1}^{m}}\mu_{i}\overline{c}_{p_{i}}\cdot\Theta_{i}+\overline{\dot{Q}}}-\dfrac{\gamma_{1}-1}{\gamma_{1}}\cdot\dfrac{\overline{R}_{m}}{\overline{c}_{p_{m}}},\label{eq1}
\end{equation}
 where $\mu_{i(j)}=\dfrac{G_{i(j)}}{G_{1}}$, $\Theta_{i}=\dfrac{T_{i}^{*}}{T_{1}^{*}}$,
$\overline{c}_{p_{m}}=\dfrac{c_{p_{m}}}{c_{p_{1}}}=\dfrac{\sum\limits _{j=1}^{l}c_{p_{j}}\theta_{j}\mathbf{\mu}_{j}}{\sum\limits _{j=1}^{l}\theta_{j}\mathbf{\mu}_{j}}$,
$\overline{c}_{v_{i}}=\dfrac{c_{v_{i}}}{c_{v_{1}}}$, $\overline{c}_{v_{m}}=\dfrac{\sum\limits _{j=1}^{l}c_{v_{m}}\theta_{j}\mu_{j}}{\sum\limits _{j=1}^{l}\theta_{j}\mu_{j}}$,
$\overline{R}_{m}=\dfrac{R_{m}}{R_{1}}=\dfrac{\sum\limits _{j=1}^{l}R_{j}\mathbf{\theta}_{j}\mathbf{\mu}_{j}}{\sum\limits _{j=1}^{l}\mathbf{\theta}_{j}\mathbf{\mu}_{j}}$,
$R_{m}=\dfrac{\sum\limits _{j=1}^{l}R_{j}G_{j}}{\sum\limits _{j=1}^{l}G_{j}}\text{, }$$\overline{\dot{Q}}=\dfrac{\dot{Q}}{c_{p_{1}}T_{1}^{*}G_{1}}$,
$\gamma_{1}=\dfrac{c_{p_{1}}}{c_{v_{1}}}$, $\overline{R}_{i}=\dfrac{R_{i}}{R_{1}}$,
$R_{i}$, $R_{m}$ -- gas constant of $i$-inlet ($i=1,2,\ldots,m$)
and gas constant of the mixture respectively FES; $c_{p_{j}},\: T_{j}^{*}$
-- heat capacity at constant pressure and total temperature of gas
$j$-outlet respectively; $T_{i}^{*},\: c_{p_{i}}\:(i=1,2,\ldots,m)$
-- total temperature and heat capacity at constant gas pressure of
$i$- inlet FES; $c_{p_{m}}$ -- heat capacity of gas at constant mixture
pressure; $G_{i}\:(i=1,2,\ldots,m)$ -- consumption of the medium through
the $i$-inlet; $G_{j}\:(j=1,2,\ldots,k)$ -- consumption through
the $j$-outlet; $a_{crit(j)}$, $V_{i(j)}$, $\lambda_{i(j)}=\dfrac{V_{i(j)}}{a_{crit(j)}}$
-- the critical speed and gas velocity at the $i$-inlet ($j$-outlet)
in the working chamber of FES and speed coefficient respectively;
$\dot{Q}$ -- non-mechanical energy (power), carried to or out of
FES.

\section{Theorem proving }

To determine the maximum evaluation of the effectiveness of energy
conversion in FES, we consider gas as ideal and compressed. In addition,
for heterogeneous gases entering the $m$-inlets of FES, we consider
the process of mixing in the working chamber of FES completed (Fig.
1).

\begin{figure}[H]
\epsfxsize=1.0\textwidth \centerline{\psfig{figure=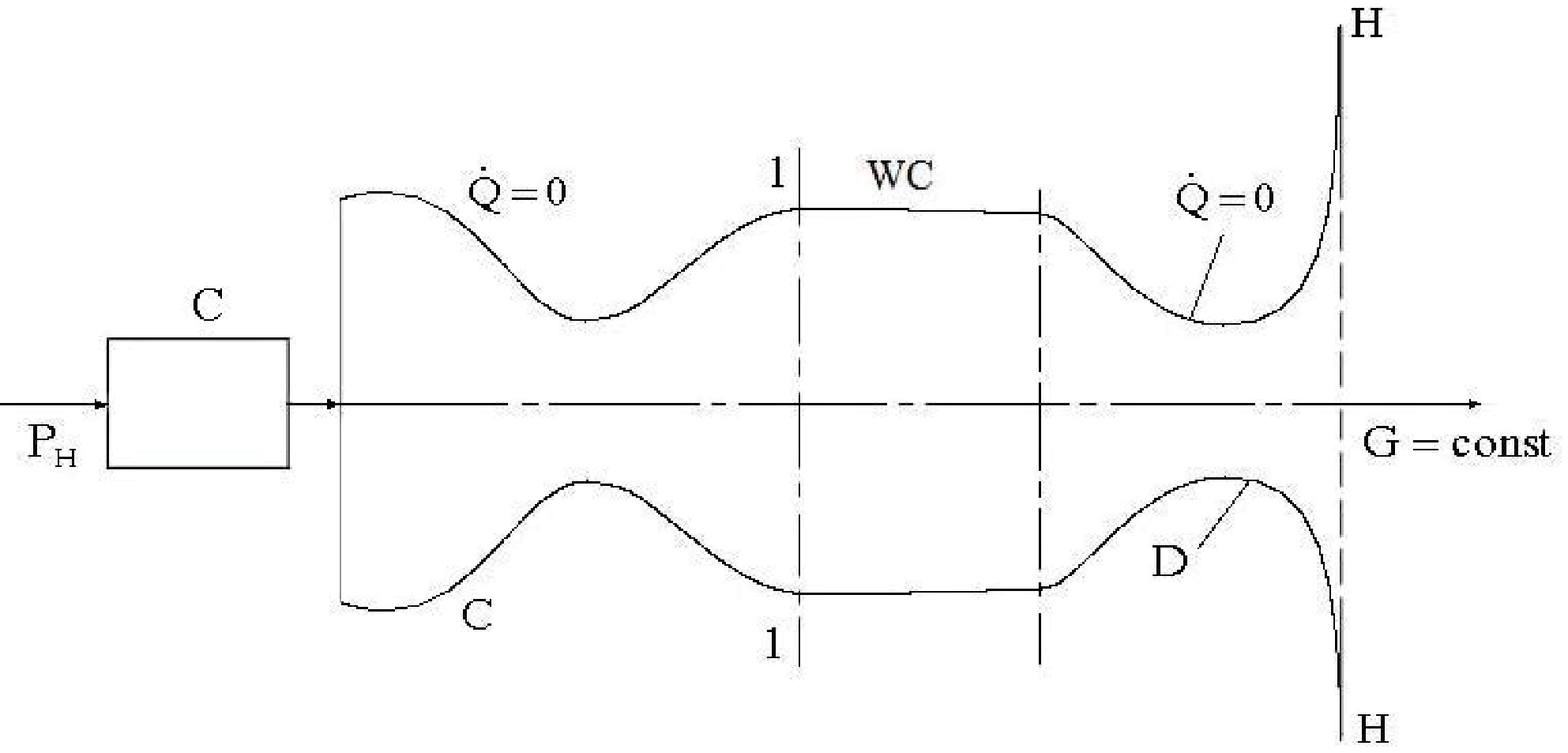,
height=7.5cm,width=10 cm}} \vspace{-0.2cm}
\caption{Fundamental diagram of the FES ($L_{tech}=0$) with one outlet ($i=1$)
and one outlet ( $j=1$), where $C$ - compressor; $WC$ -- working
chamber; $D$ -- diffuser; $p_{H}$, $p_{K}$ -- pressure at the inlet
and outlet of the compressor correspondingly, and C - nozzle. Section
1 -- inlet into the working chamber, H -- outlet of FES. }
\end{figure}

Total mechanical energy (power) of flow at the inlet to the FES and
the energy (power) delivered from outside are
\[
N_{ex}=\sum\limits _{i=1}^{m}\left(\dfrac{p_{i}}{\rho_{i}}+\dfrac{\text{v}_{i}^{2}}{2}\right)G_{i}+\dot{Q}=\sum\limits _{i=1}^{m}G_{i}R_{i}T_{i}^{\ast}\left[\left(1-\dfrac{\gamma_{i}-1}{\gamma_{i}+1}\lambda_{i}^{2}+\dfrac{\gamma_{i}}{\gamma_{i}+1}\lambda_{i}^{2}\right)\right]+\dot{Q}
\]
\begin{equation}
=\sum\limits _{i=1}^{m}G_{i}R_{i}T_{i}^{\ast}\left(1+\dfrac{\lambda_{i}^{2}}{\gamma_{i}+1}\right)+\dot{Q},\label{eq2}
\end{equation}
 where $p_{i}$, $\rho_{i}\text{,}$ $v_{i}$ -- static pressure,
density and velocity at the $i$-inlet to the heat engine, correspondingly.
The total energy (power) delivered to FES is
\begin{equation}
N_{\Sigma}=\sum\limits _{i=1}^{m}G_{i}i_{i}^{*}+\dot{Q}=\sum\limits _{i=1}^{m}c_{p_{i}}T_{i}^{*}G_{i}+\dot{Q}.\label{eq3}
\end{equation}
The total energy (power) of gas flows at the outlet of FES diffuser
is determined by
\begin{equation}
N_{out}=R_{m}\sum\limits _{j=1}^{k}T_{j}^{\ast}G_{j}\left(1+\dfrac{1}{\gamma_{j}+1}\lambda_{j}^{2}\right).\label{eq4}
\end{equation}
 For the stationary case the nondimensional equations of continuity
and energy in the absence of technical work (L$_{tech}$= 0) take
the form
\begin{equation}
1+\sum\limits _{i=2}^{m}\mu_{i}=\sum\limits _{j=1}^{\kappa}\mu_{j}, \,\, 1+\sum\limits _{i=2}^{m}\overline{c}_{p_{i}}\Theta_{i}\mu_{i}+\overline{\dot{Q}}=\overline{c}_{p_{m}}\sum\limits _{j=1}^{n}\Theta_{m}\mu_{j}.\label{eq5}
\end{equation}

We define the limit of the relative share of mechanical power flow
utilized in FES taking into account (\ref{eq5}) at $\lambda_{j}\rightarrow0$ which
corresponds to an infinite broadening of the diffuser $S\rightarrow\infty$
(Fig. 1).

\[
\Delta\overline{N}_{max}=\underset{\lambda_{j}\rightarrow0}{\lim\Delta\overline{N}}
\]

\[
=\dfrac{\displaystyle{\sum\limits _{i=1}^{m}}\mu_{i}\overline{R}_{i}\Theta_{i}\left(1+\dfrac{\lambda_{i}^{2}}{\gamma_{i}+1}\right)\dfrac{\gamma_{1}-1}
{\gamma_{1}}+\overline{\dot{Q}}-\dfrac{\overline{R}_{m}}{\overline{c}_{p_{m}}}\left[\displaystyle{\sum\limits _{i=1}^{m}}\overline{c}_{p_{i}}\Theta_{i}\mu_{i}+\overline{\dot{Q}}\right]\dfrac{\gamma_{1}-1}{\gamma_{1}}}{\displaystyle{\sum\limits _{i=1}^{m}}\overline{c}_{p_i}\Theta_{i}\mu_{i}+\overline{\dot{Q}}}
\]
\begin{equation}
=\dfrac{\displaystyle{\sum\limits _{i=1}^{m}}\mu_{i}\overline{R}_{i}\Theta_{i}\left(1+\dfrac{\lambda_{i}^{2}}{\gamma_{i}+1}\right)\dfrac{\gamma_{1}-1}{\gamma_{1}}+\overline{\dot{Q}}}{\displaystyle
{\sum\limits _{i=1}^{m}}\overline{c}_{p_{i}}\Theta_{i}\mu_{i}+\overline{\dot{Q}}}-\dfrac{\overline{R}_{m}}{\overline{c}_{p_{m}}}\dfrac{\gamma_{1}-1}{\gamma_{1}}.\label{eq6}
\end{equation}
Since the rate of flow at the outlet of FES is nonzero, then the
efficiency of energy transformation in ideal FES is less $\eta_{G>0}^{ideal}<\Delta\overline{N}_{max}$
Q.E.D which was to be proved.

\section{Consequences }

A global maximum of utilized mechanical power flow, as well as non-mechanical
power carried out outside the working chamber in the FES equals
\begin{equation}
\Delta\overline{N}_{max}^{Global}=\lim_{\substack{\lambda_{i}\rightarrow0 \\ \lambda_{\text{j}}\rightarrow\lambda_{max}}}\Delta\overline{N}_{max}=\dfrac{\displaystyle{\sum\limits _{i=1}^{m}}\mu_{i}\overline{R}_{i}\Theta_{i}\dfrac{\gamma_{i}}{\gamma_{i}-1}\cdot\dfrac{\gamma_{1}-1}{\gamma_{1}}+\overline{\dot{Q}}}{\displaystyle{\sum\limits _{i=1}^{m}}\overline{c}_{p_{i}}\Theta_{i}\mu_{i}+\overline{\dot{Q}}}-\dfrac{\gamma_{1}-1}{\gamma_{1}}\cdot\dfrac{\overline{R}_{m}}{\overline{c}_{p_{m}}}.\label{eq7}
\end{equation}

In case of temperatures and the same name gases consumption being
equal at $m$-inlets into the working chamber of FES equation (\ref{eq6})
takes the form
\begin{equation}
\Delta\overline{N}_{max}=\dfrac{1}{\gamma},\textrm{ where }\gamma=\dfrac{c_{p}}{c_{v}}.\label{eq8}
\end{equation}

Expression (\ref{eq7}) for the same name gases with significant contribution
of energy in FES (\ref{eq8}) has the form $\overline{\dot{Q}}\rightarrow\infty$
(\ref{eq8}).

Ratio of efficiency of energy conversion in an ideal FES can not exceed
the efficiency factor of ideal Carnot cycle.

From (\ref{eq1}) for the same name gases at $m$-inlets at open-circuit
$\overline{\dot{Q}}\rightarrow 0$ we obtain:
\begin{equation}
\Delta\overline{N}_{max}=\dfrac{\dfrac{1}{\gamma}\eta_{Carnot}^{ideal}+\dfrac{1}{\gamma}\overline{\dot{Q}}}{1+\overline{\dot{Q}}},\label{eq9}
\end{equation}
 where
\[
\eta_{Carnot}=1-\dfrac{T_{1}}{T_{1}^{*}}=\dfrac{\gamma-1}{\gamma+1}\lambda_{1}^{2}.
\]
 From (\ref{eq9}) it follows that the ratio of energy conversion
efficiency of FES $\eta_{max}^{ideal}$ increases with gas velocity
at the inlet. In addition, the formula (8) gives the relation of the
three theorems: the proved theorem, and theorems of S. Carnot [1]
and W.~Nernst [2]. Ratio of energy conversion efficiency in a
gas machine ($\eta_{G>0}^{ideal},\:\dot{Q}=0$) will be ã times less
than gas dynamic efficiency of Carnot cycle ($\dfrac{1}{\gamma}\eta_{Carnot}^{ideal}$).
At the same time even for ideal gas due to inaccessibility of absolute
zero $T_{2}>0$ (Nernst theorem), Carnot efficiency is less than one,
therefore, for the ratio of energy conversion efficiency in a flow
machine we get
\begin{equation}
\eta_{G>0}^{ideal}=\dfrac{1}{\gamma}\eta_{Carnot}^{ideal}<\dfrac{1}{\gamma}.\label{eq10}
\end{equation}

\section{Discussion of results}

Theorem gives a more strongly worded II law of thermodynamics for
the selected class of flow energy systems: only part of the energy
of gas flow in an ideal FES can be converted into useful work (the
effect). For the case of the same name gas at m-inlets of FES, this
part will be $\dfrac{1}{\gamma}$ of the total energy of the gas flow.

Let us consider non mixing gases at the gas flow machine outlet. In
this case, the estimate of the marginal efficiency of energy conversion
in the gas flow machine is just the summation of solution (8) for
$m$-inputs and $n$-outputs
\begin{equation}
\Delta\overline{N}_{i}=\dfrac{\dfrac{1}{\gamma_{i}}\eta_{i}+\dfrac{1}{\gamma_{i}}
\cdot\overline{\dot{Q}}_{i}}{1+\overline{\dot{Q}}_{i}}\:\Rightarrow\:\Delta\overline{N}_{i}=\dfrac{\Delta N_{i}}{\displaystyle{\sum_{i=1}^{m}}
c_{p_i}T_{i}^{\ast}G_{i}+\dot{Q}_{i}},\label{eq11}
\end{equation}
 hence
\[
\Delta N_{i}=\Delta\overline{N}_{i}c_{p_i}T_{i}^{\ast}G_{i}(1+\overline{\dot{Q}}_{i}).
\]
 The total value of utilized mechanical energy flow at the $m$-inputs
is equal to
\[
\Delta N_{\Sigma}=\sum\limits _{i=1}^{m}c_{p_i}T_{i}^{\ast}\overline{\dot{Q}}_{i}(1+\overline{\dot{Q}}_{i})\Delta\overline{N}_{i},
\]
 hence
\[
\Delta\overline{N}_{\Sigma}=\dfrac{\Delta N_{\Sigma}}{N_{\Sigma}}=\dfrac{{\displaystyle\sum\limits _{i=1}^{m}}c_{p_i}T_{i}^{\ast}G_{i}(1+\overline{\dot{Q}}_{i})\Delta\overline{N}_{i}}{i_{\Sigma}^{\ast}+\dot{Q}_{\Sigma}}
\]
\begin{equation}
=\dfrac{{\displaystyle\sum\limits _{i=1}^{m}}i_{i}^{\ast}(1+\overline{\dot{Q}}_{i})\Delta\overline{N}_{i}}{i_{\Sigma}^{\ast}( 1+\overline{\dot{Q}}_i)}.\label{eq12}
\end{equation}
 In accordance with the average theorem we obtain
\begin{equation}
\eta_{G>0}^{ideal}=\Delta\overline{N}_{\Sigma}=\langle\Delta\overline{N}_{i}\rangle.\label{eq13}
\end{equation}
 For open-circuit ( $\overline{\dot{Q}}=O$) and non mixing gas flows
we get
\begin{equation}
\eta_{G>0}^{ideal}=\langle\dfrac{1}{\gamma_{i}}\dfrac{\gamma_{i}-1}{\gamma_{i}+1}\lambda_{i}^{2}\rangle=\langle\dfrac{1}{\gamma_{i}}\eta_{i}^{Carnot}\rangle\label{eq14}
\end{equation}
 and for maximum input speed ($\lambda_{i}=\sqrt{\dfrac{\gamma_{i}+1}{\gamma_{i}-1}}$)
we get
\begin{equation}
\eta_{G>0}^{ideal}=\langle\dfrac{1}{\gamma_{i}}\rangle.\label{eq15}
\end{equation}

As opposed to the consequences of Carnot theorem, the proved theorem
(\ref{eq10}) implies that the efficiency of energy conversion in
an ideal FES depends on the properties of the working mass ($\gamma=c_{p}/c_{v}$).
In this case the conditional cycle in FES is essentially irreversible
(Fig. 2, 3).

The ideal conditional cycle in $P-V$ coordinates presented in Fig. 2,
for the case of open-circuit $\overline{\dot{Q}}\rightarrow0$, consists
of one ideal isotherm ($H-K$), two ideal adiabats ($K-1$, $1'-H$) and one
percussive adiabat of Hugoniot ($1-1'$).

It should be noted that in the supersonic flow regime ( $\lambda_{1}>1$)
there is always a shock wave, since the flow regime in the nozzle
is off calculation. Especially clear the distinction of conditional
cycle in FES from the Carnot cycle can be seen in $T-S$ coordinates
(Fig. 3). The difference between areas in Carnot cycle and PES (curvilinear
triangle $1-1'-2-1$) represents the energy required for pumping the
environment in FES.

The crosshatched area of $HK-1-1'-H$, referred to the total area of
the curve $KH-1'-1$, represents the relative share of lost or utilized
energy in FES: \hspace{5mm}
\begin{equation}
\eta_{G>0}^{ideal}=\Delta\overline{N}=1-\dfrac{\int\limits _{V_{H}}^{V_{K}}PdV}{\int\limits _{V_{R}}^{V_{1}}PdV+\int\limits _{V_{1}}^{V_{1^{\prime}}}PdV+\int\limits _{V_{1^{\prime}}}^{V_{H}}PdV}.\label{eq16}
\end{equation}
 According to the theorem, this value can not exceed the corresponding
maximum value determined by the formula
\[
\eta_{G>0}^{ideal}\leq\eta_{max}^{ideal}=\Delta\overline{N}_{max}\,.
\]
 It should be noted that in $P-V$ coordinates in general case the trajectory
of the Hugoniot shock adiabat is not defined, there are only the initial
( $P_{1}$, $V_{1}$) and final ( $P_{1}^{\prime}$, $V_{1}^{\prime}$)
values of the trajectory. For the case of confluent shock wave in
L.D.Landau's work \cite{key-3} there obtained a solution to the problem
of the curvature trajectory of Hugoniot adiabat ($\partial^{2}V/\partial P^{2}>0$).
However, extrapolation of given result in the sphere of strong shock
waves is not argued, and the concavity of the curve ($\partial^{2}P/\partial^{2}V>0$)
is taken as a hypothesis.

To determine the shape of the curve at the section ($1-1'$) as a criterion
value $\eta_{max}^{ideal}\leq\Delta\overline{N}$ is used while varying
the pressure at outlet from the system ($P_{H}$) ) and fixed value
$P_{K}=P_{K}^{\ast}$. In the first approximation at the area $1-1'$,
a linear relationship between P and V was used. In this case, the
area under the curve $1-1'$ is defined as the area of a trapezoid.

Numerical calculations have shown that up to values of the velocity
coefficient at the entrance to the chamber $\lambda_{1}=2.307$ ($\gamma=1.4$),
criterion estimation obtained in the theorem will be satisfied on
condition that
\begin{equation}
\int\limits _{1}^{1'}PdV< \dfrac{(p_{1}+p_{1'})}{2}(V_{1}-V_{1'}), \label{eq17}
\end{equation}
 i.e. the conditional trajectory $1-1'$ should be concave (Fig. 2).
\begin{figure}[p]
\epsfxsize=1.0\textwidth \centerline{\psfig{figure=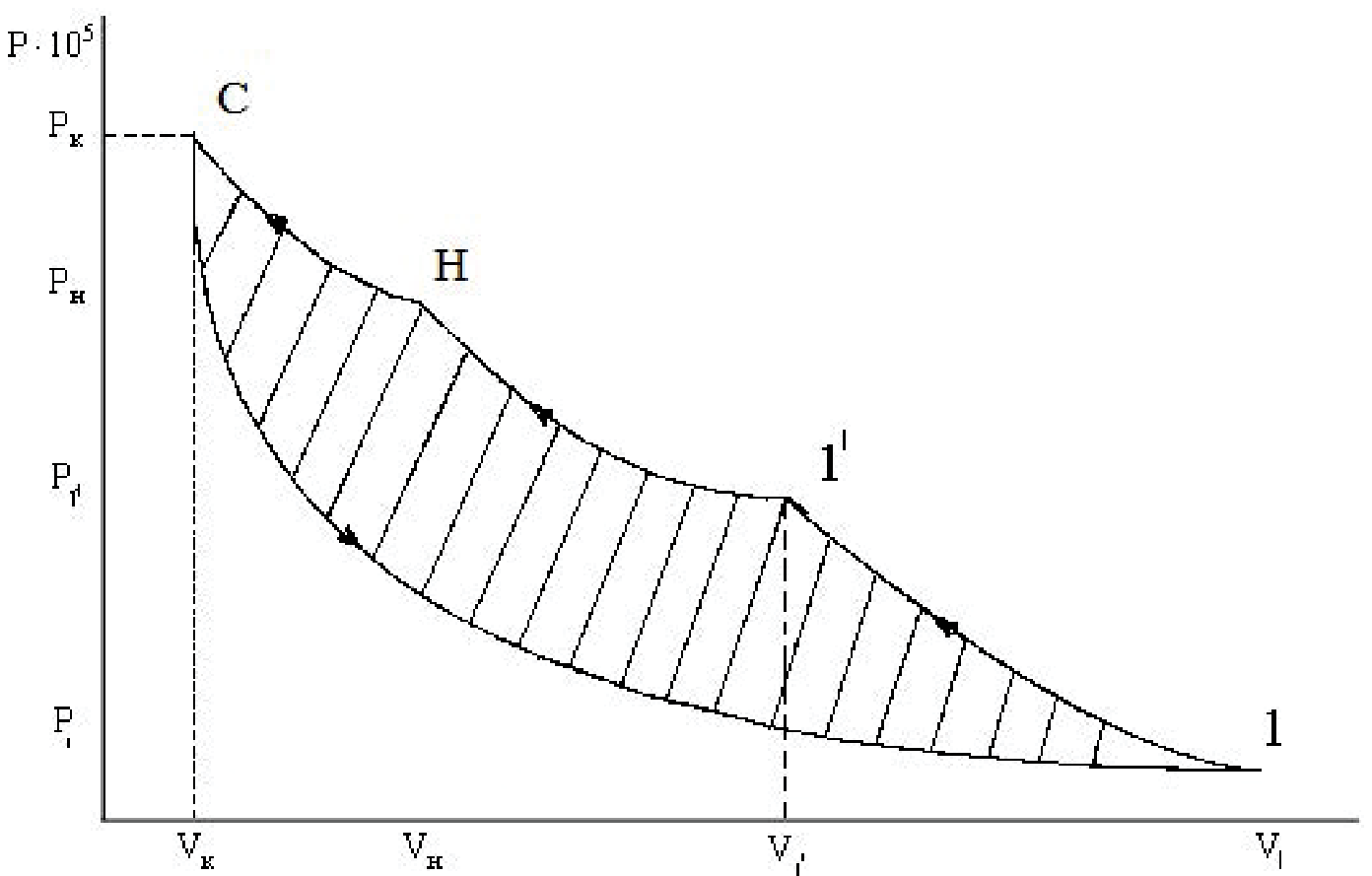,
height=7.5cm,width=10cm}} \vspace{-0.2cm}

 \caption{
Conditional limit cycle of open-circuit $G''C$
($\overline{\dot{L}}_{tech}=0$). $H-C$ -- isothermal compression, for example, in
the compressor; $K-1$ --adiabatic expansion in the nozzle; $1-1'$ --
Hugoniot shock adiabat; $1'-H$ -- adiabatic compression in the diffuser. }

\end{figure}

\begin{figure}[p]

\epsfxsize=1.0\textwidth \centerline{\psfig{figure=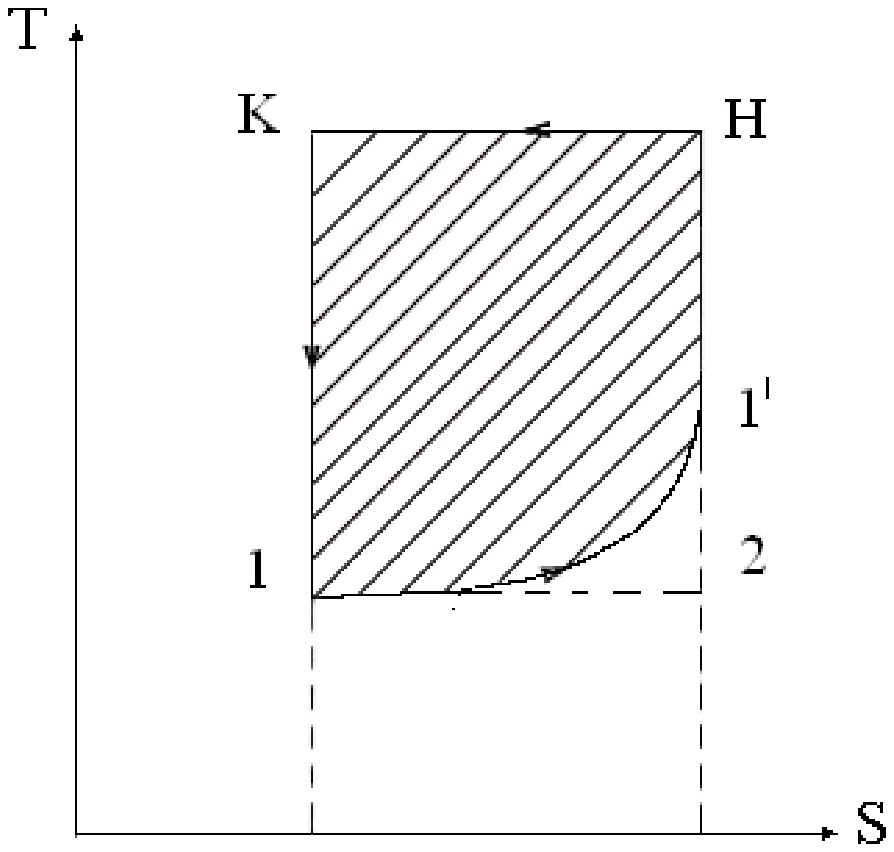,
height=7.5cm,width=7.4cm}} \vspace{-0.2cm}

 \caption{Limit cycle FES in $T-S$ coordinates. $H-K$ -- isothermal compression,
for example, in the compressor, $K-1$ -- adiabatic expansion in the
nozzle; $1-1'$ -- shock wave (Hugoniot percussive adiabat); $1'-H$ --
adiabatic compression in the diffuser. $K-1-2-H-K$ -- Carnot cycle.}

\end{figure}

Thus, the theorem without any additional hypotheses makes it possible
to determine the shape of the conditional process trajectory at the
site of the shock adiabat in P-V coordinates for the case of strong
shock waves.

The table presents the classification of energy systems (machines)
depending upon the directions of energy utilization used in them.

\begin{table}[p]
\caption{Classification of energy systems  according
to the method of the working body total energy conversion}
\begin{tabular}{|p{1in}|p{1.4in}|p{1.3in}|p{1.3in}|}

\hline
\textbf{\small Class FES} & \textbf{\small I} & \textbf{\small II} & \textbf{\small III}\tabularnewline
\hline
{\small Consumption of gaseous working mass } & {\small $G=0$} & {\small $G>0$}{\small \par}

{\small $G\rightarrow G_{max}$} & {\small $G>0$}{\small \par}

{\small $\left(G\rightarrow G_{max}\right)$ }\tabularnewline
\hline
{\small Gas velocity at the outlet of FES } & {\small $V_{out}\ll a_{sound}$ } & {\small $\left(V_{out}\ge  a_{sound}\right)$ } & {\small $V_{out}\to0$}{\small \par}{\small $\left(S_{out}\to\infty\right)$ }\tabularnewline
\hline
{\small Direction of conversion of gas flow total energy } & {\small Total (internal) energy is converted into mechanical work
$E_{tot}=E_{int}\leftrightarrow A_{mec}$ } & {\small Total energy is converted into kinetic energy }{\small \par}

{\small $E_{tot}\rightarrow E_{kin}$} & {\small Total energy is converted into potential energy of pressure}{\small \par}

{\small $E_{tot}\rightarrow E_{pot}$}\tabularnewline
\hline
{\small Efficiency } & {\small $$\eta_{Carnot}^{{ideal}}=1-\dfrac{T_{1}}{T_{k}}$$ $$\eta_{Stirling}
{\rm ~ets.}$$} & {\small $$\eta=1-\left(\dfrac{p_{0}}{p_{1}}\right)^{\dfrac{n-1}{n}}$$ $$\lim_{p_{1}\rightarrow\infty}\eta\to\eta_{Carnot}^{{ideal}}$$} & {\small $$L_{tech}=0$$
$$\overline{Q}\to 0 $$  $$\eta_{max}^{ideal}=\dfrac{1}{\gamma}\eta_{Carnot}^{ideal}$$}\tabularnewline
\hline
{\small Cycle of heat machine} & {\small Carnot cycle Stirling Otto, Diesel, etc. } & {\small Cycle Brighton, Humphrey, etc. } & {\small Cycle PES }\tabularnewline
\hline
\end{tabular}
\end{table}

If the total energy is converted into mechanical work, then we deal
with machines running on the Otto, Diesel, Rankine, Stirling cycles,
etc., then a limit cycle in this direction of energy transformation
is the Carnot cycle. The opposite direction is a cycle of refrigerating
machines. When converting the total energy into kinetic energy of
the gas flow at the outlet of FES, we deal with jet and air-breathing
jet engines and, accordingly, with the conventional cycles of Brayton,
Humphrey, pulsating air-breathing jet engine, where the limit cycle
is also Carnot\textquoteright{}s. In the case of energy flow conversion
in FES in the potential energy of pressure the limit cycle will be
cycle FES (Fig. 2, 3).

From Fig. 4, which shows a spatial illustration of the given classification
of energy systems, it follows that the proved theorem that has given
the limit estimation of energy utilization in FES, allowed to move
from flat space heat systems (machines) to three-dimensional, where
the coordinates of any vector $\overrightarrow{M}$ represent the
characteristics of a some combined flow energy system.

\begin{figure}[p]
\epsfxsize=1.0\textwidth \centerline{\psfig{figure=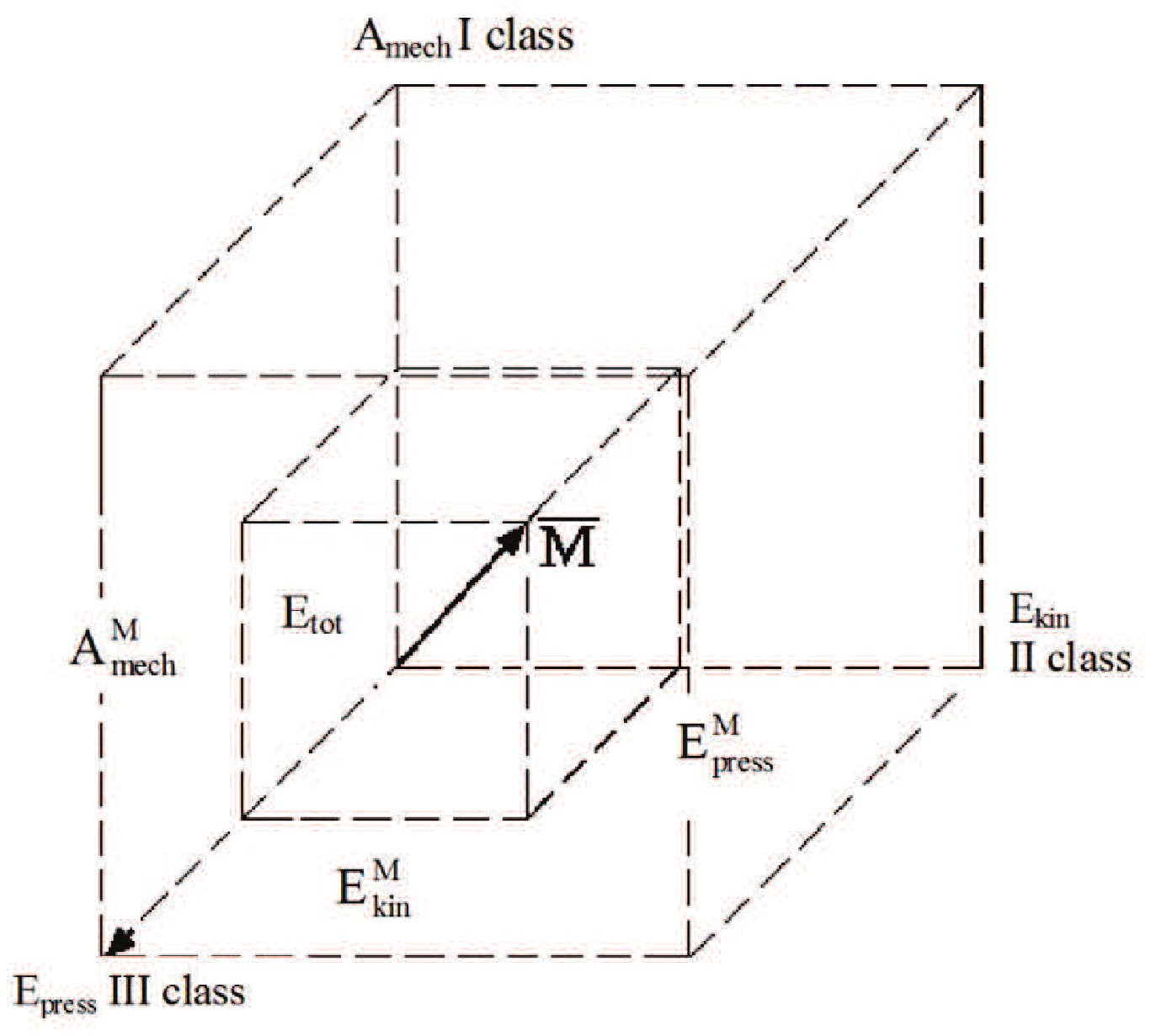,
height=8cm,width=10cm}} \vspace{-0.2cm}

\caption{The space of flow energy systems. }

\end{figure}

The effectiveness of energy systems FES can be defined by the indicator
of flow process quality I, which equals to the ratio of useful utilized
energy to the maximum possible energy that can be utilized in this
device
\begin{equation}
I=\dfrac{E_{ut}}{\Delta N_{max}},\label{eq18}
\end{equation}
where $E_{ut}$ - utilized (useful) energy in FES.

Full effectiveness of this class machines is defined as follows
\begin{equation}
\eta^{\Sigma}=\prod_{i=1}^{N}\eta_{i},\: i=1,\ldots,n.\label{eq19}
\end{equation}

For example, for a gas laser (with pumping gas medium) electrically
pumped full effectiveness (or efficiency) can be written as
\begin{equation}
\eta_{laser}=\eta_{gas}\eta_{serv}\eta_{eo}\eta_{quant},\label{eq20}
\end{equation}
where $\eta_{gas}=I\cdot\eta_{max}^{ideal}$; $\eta_{serv}$ -- is
the effectiveness of the gas path, the effectiveness of service laser;
$\eta_{eo}$, $\eta_{quant}$ -- is electro-optical and quantum efficiency
of gas laser.

So for $CO_{2}$ laser $\eta_{quant}\sim0.4$, and for $CO$ laser
$\eta_{quant}\sim0.8$ respectively, i.e. full efficiency of laser
system, even in the ideal case ($\eta_{eo}=\eta_{serv}=1$) can not
exceed for $CO_{2}$ and $CO$ lasers, correspondingly, values
\[
\eta_{CO_{2}}^{\Sigma,ideal}=\dfrac{1}{\gamma_{m}}\eta_{quant}^{CO_{2}}\approx0.25,\:\eta_{CO}^{\Sigma,ideal}=\dfrac{1}{\gamma_{m}}\eta_{quant}^{CO}\approx0.5,\:\gamma_{m}=c_{p_{m}}/c_{v_{cm}}.
\]

The results of the theorem can be used to estimate the energy release
in astrophysical thermodynamic systems (in disks of close stars systems).
In these energy systems the conditions of the theorem are ideally
suited - the technical work is absent ($L_{tex}=0$), gas velocity
at the outlet from the system is zero $\lambda_{1}=0$. For the case
of weak gravitational fields and small relativistic gas velocity $v<c$,
we obtain on the basis of Consequence  1 of the theorem
($\overline{Q}\rightarrow\infty$), estimation of share energy radiation
in close stars from $Gc^{2}$ recieve $E_{rel}<\dfrac{1}{\gamma}Gc^{2}$.

\vskip 0.5cm
\noindent
{\large \bf Acknowledgements}
\vskip 0.2cm
\noindent
I am very grateful to Prof. A.I. Leontiev and Prof. V.C. Shakhov for the useful advice concerning the theorem proving, to Dr. A.P.Zubarev and T.A. Bazhenova for assistance in preparing the article.

\end{document}